\documentclass[iop,apj]{emulateapj}

\newcommand{\angstrom}{\textup{\AA}}
\newcommand{\sunrise}{\textsc{Sunrise}}
\newcommand{\carcsec}{$\mbox{.\hspace{-0.5ex}}^{\prime\prime}$}

\shortauthors{Kahil et al.}

\usepackage{multirow}
\usepackage{natbib}
\usepackage{amsmath}
\usepackage{color}
\begin{document}

\title{Brightness of solar magnetic elements as a function of magnetic flux at high spatial resolution}

\author{\textsc{
F.~Kahil,$^{1}$
T.~L.~Riethm\"uller,$^{1}$
\& S.~K.~Solanki,$^{1,2}$
}}
\affil{
$^{1}$Max-Planck-Institut f\"ur Sonnensystemforschung, Justus-von-Liebig-Weg 3, 37077 G\"ottingen, Germany; kahil@mps.mpg.de\\
$^{2}$School of Space Research, Kyung Hee University, Yongin, Gyeonggi, 446-701, Republic of Korea
}

\begin{abstract}
We investigate the relationship between the photospheric magnetic field of small-scale magnetic elements in the quiet Sun (QS) at disc centre, and the brightness at 214 nm, 300 nm, 313 nm, 388 nm, 397 nm, and at 525.02 nm. To this end we analysed spectropolarimetric and imaging time series acquired simultaneously by the IMaX magnetograph and the SuFI filter imager on-board the balloon-borne observatory $\sunrise$ during its first science flight in 2009, with high spatial and temporal resolution.

We find a clear dependence of the contrast in the near ultraviolet (NUV) and the visible on the line-of-sight component of the magnetic field, $B_{\rm LOS}$, which is best described by a logarithmic model. This function represents well the relationship between the Ca\,{\sc ii} H-line emission and $B_{\rm LOS}$, and works better than a power-law fit adopted by previous studies. This, along with the high contrast reached at these wavelengths, will help with determining the contribution of small-scale elements in the QS to the irradiance changes for wavelengths below 388 nm. At all wavelengths including the continuum at 525.40 nm the intensity contrast does not decrease with increasing $B_{\rm LOS}$. This result also strongly supports that $\sunrise$ has resolved small strong magnetic field elements in the internetwork, resulting in constant contrasts for large magnetic fields in our continuum contrast at 525.40 nm vs. $B_{\rm LOS}$  scatterplot, unlike the turnover obtained in previous observational studies. This turnover is due to the intermixing of the bright magnetic features with the dark intergranular lanes surrounding them.

\end{abstract}


\section{Introduction} 
Small scale magnetic elements or magnetic flux concentrations are described by flux tubes, often with kG field strengths, located in intergranular downflow lanes \citep{solanki93}. 
Studying the intensity contrast of magnetic elements relative to the QS in the continuum, and line core of spectral lines, is of importance, because it provides information about their thermal structure. Due to their enhanced brightness, particularly in the cores of spectral lines \citep[]{title92, yeo13} and in the UV \citep{tino10}, these elements are believed to contribute to the variation of the solar irradiance, especially on solar cycle time scales \citep{f_l88,fligge2000, krivova03, yeo14}. The contrast in the visible, and in the UV spectral ranges contributes by 30\% and 60\% respectively, to the variation of the total solar irradiance (TSI) between minimum and maximum activity \citep{krivova06} and spectral lines contribute to a large part of this variation \citep[]{livi88, shapiro15}.
The magnetic flux in magnetic elements is also believed to be responsible for the structuring and heating of the chromosphere and corona. The relationship with chromospheric heating is indicated by the strong relationship between excess brightening in the core of the Ca\,{\sc ii} K-line and the photospheric magnetic flux \citep[e.g.,][]{sku75,sch89,louki09}.  \\ 

Several studies have been carried out to investigate how the brightness at given wavelength bands depends on the photospheric magnetic field at disc centre, and different results have been obtained. \citet{title92} and \citet{topka92} made a pixel-by-pixel comparison of the longitudinal magnetogram signal and the continumm intensity at 676.8, 525, 557.6, and 630.2 nm, using simultaneous continuum filtergrams, in addition to line centre filtergrams at 676.8 nm in \citet{title92}, and magnetograms of magnetic features in active regions at disc centre, acquired by the 50 cm Swedish Solar Vacuum Telescope with a 0.3$^{\prime\prime}$ resolution. They obtained negative contrasts in the continuum (darker than the average QS) for all magnetogram signals (as long as the observations were carried out almost exactly at solar disc centre), and an increase in the line core brightness with the magnetogram signal until 600\,G followed by a monotonic decline.

\citet{lawrence93} applied the same method at roughly the same spatial resolution as \citet{topka92} to quiet-Sun network data. Their scatterplot showed positive contrast values (brighter than the average QS) for magnetic fields larger than 200\,G, until nearly 500\,G, and negative contrasts for higher magnetic fields.
\citet{ortiz02} used low resolution (4$^{\prime\prime}$), but simultaneous magnetograms and full-disc continuum intensity images of the Ni\,{\sc i} 6768\, \AA{} absorption line, recorded by the Michelson Doppler Imager (MDI) on-board the Solar and Heliospheric Observatory (SoHO). They found that the contrast close to disc centre (at $\mu \approx 0.96$) initially increases slightly with the magnetic field, before decreasing again for larger fields.
\citet{kobel11} made the same pixel-by-pixel study of the continuum contrast at 630.2 nm with the longitudinal magnetic field in the quiet-Sun network and in active region (AR) plage near disc centre, using data from the Solar Optical Telecope on-board Hinode (0.3$^{\prime\prime}$ spatial resolution). They found that for both the QS and AR, the contrast initially decreases and then increases for weak fields, until reaching a peak at $\approx$ 700\,G, to decrease again for stronger fields, even when the pores in their AR fields of view (FOVs) were expilicitly removed from their analysis.
\\ They explained the initial, rapid decrease to have a convective cause, with the bright granules harbouring weaker fields than the intergranular lanes. The following increase in contrast is due to the magnetic elements being brighter than the average QS. To explain the final decrease at high field strengths, they argued that due to their limited spatial resolution (0.3$^{\prime\prime}$ for Hinode/SP), many flux tubes were not resolved, and therefore, the field strength of small bright elements is attenuated, while those of bigger structures (such as micropores), which are darker than the mean QS, were weakly affected by the finite resolution, so that the average contrast showed a peak at intermediate field strengths, and decreased at higher strengths.

At a constant spatial resolution of 1$^{\prime\prime}$ achieved by the Helioseismic and Magnetic Imager (HMI) on-board the Solar Dynamics Observatory (SDO), \citet{yeo13} studied the dependence of the continuum and line core contrasts in  the Fe\,{\sc i} line at 6173\, \AA{}, of network and faculae regions on disc position and magnetogram signal, using simultaneous full-disc magnetograms and intensity images. For a quiet-Sun region at disc centre ($\mu > 0.94$), their scatterplots of the continuum intensity against $<B>/\mu$ exhibited a peak at $<B>/\mu \sim$ 200\,G.  \\

\citet{roh11} simulated a plage region using the MURaM code \citep{vog05}. They studied the relation between the continuum contrast at 630.2 nm and the vertical magnetic field both, at the original MURaM resolution and at the resolution of telescopes with 1.0 m and 0.5 m apertures. For the original resolution, the contrast monotonically increased with increasing field strength, confirming the expectations of a thin flux-tube model \citep{spruit76}, which predicts that for higher field strengths, the flux tubes get more evacuated, which leads to lateral inflow of heat from the hot walls of the evacuated flux concentrations, and therefore an increase in brightness as a result of the optical depth surface depression, which allows deeper layers to be seen. According to \citet{roh11}, at the resolution of a  1.0 m telescope the average simulated contrast saturates for stronger fields, while it shows a turnover at the resolution corresponding to 0.5 m. This points to a non-trivial effect of finite spatial resolution on the relation between continuum brightness, and photospheric magnetic field.
\\

In contrast to continuum radiation, spectral lines display a relatively monotonic increase in rest intensity with magnetic flux. Of particular importance is the Ca\,{\sc ii} H line, due to its formation in the chromosphere. \citet{frazier71} showed by using simultaneous observations of the calcium network and photospheric magnetic field that the line core of Fe\,{\sc i} at 5250.2 \AA{} and the Ca\,{\sc ii} K emission increases with the magnetic field until his limit of about 500\,G. \citet{sch89} carried out a quantitative study of the relationship between the Ca\,{\sc ii} K emission and magnetic flux density in an active region (outside sunspots). After subtracting the basal flux (the non-magnetic contribution of the chromospheric emission) they found that the relation follows a power law, with an exponent of 0.6.
\citet{ortiz05} found the same relation with a power-law exponent of 0.66, for a quiet-Sun region at disc centre, while \citet{rezai07} reported a value of 0.2 (including the internetwork), and 0.4--0.5 for the network, with a strong dependence of the power exponent on the magnetic field threshold.
\citet{louki09} in their study of the correlation between emissions at different chromospheric heights with the photospheric magnetic field in a quiet-Sun region close to the disc centre, found an exponent of 0.31 for Ca\,{\sc ii} K.\\

Here, we study the contrast at a number of wavelengths in the NUV (between 214 nm and 397 nm) and the visible (around 525 nm) in the QS close to solar disc centre. We employ high-resolution, seeing-free measurements of both, the intensity and the magnetic field obtained with the $\sunrise$ balloon-borne observatory. \\
The structure of the rest of the paper is as follows. In Sec. 2, we describe the data used for this analysis, in addition to presenting the detailed data reduction steps. In Sec. 3, we present our results, and compare them to the literature. In Sec. 4, we summarize and discuss the results.

\section{Observations and Data Preparation}
\subsection{IMaX and SuFI data}
We use a time series recorded during the first $\sunrise$  flight on 2009 June 9, between 14:22 and 15:00 UT. $\sunrise$ is composed of a telescope with a 1.0 m diameter main mirror mounted on a gondola with two post-focus instruments \citep{solanki010,barthol11}. It carries out its observations hanging from a stratospheric balloon. The images are stabilized against small-scale motions by a tip-tilt mirror placed in the light distribution unit connected to a correlation tracker and wavefront sensor \citep{berk11, gand11}. $\sunrise$ carried two instruments, a UV imager (SuFI) and a magnetograph (IMaX).\\

The Imaging Magnetograph eXperiment \citep[IMaX;][]{mart11} acquired the anaylsed spectropolarimetric data by scanning the photospheric Fe\,{\sc i} line at 5250.2\, \AA{} (Land{\'e} factor $g$ = 3) at five wavelengths positions (4 within the line at $-80$, $-40$, $+40$, $+80$\, m\AA{} and one in the continuum at $+227$\, m\AA{} from the line centre), with a spectral resolution of 85\, m\AA{}, and measuring the full Stokes vector $(I,Q,U,V)$ at each wavelength position. The total cadence for the V5-6 observing mode (V for vector mode with 5 scan positions and 6 accumulations of 250 ms each) was 33\,s. We use the phase-diversity reconstructed data with a noise level of 3$\times$10$^{-3} I_c$, and achieving a spatial resolution of $0.15^{\prime\prime}-0.18^{\prime\prime}$ (see \citet{mart11} for more details on the instrument, data reduction and data properties).

For the NUV observations, we use the data acquired by the $\sunrise$ Filter Imager \citep[SuFI;][]{gand11} quasi-simultaneously with IMaX, in the spectral regions 214 nm, 300 nm, 313 (OH-band) nm, 388 nm (CN-band), and 397 nm (core of Ca\,{\sc ii} H), at a bandwidth of 10 nm, 5 nm, 1.2 nm, 0.8 nm, and 0.18 nm, respectively. The cadence of the SuFI data for a given wavelength was 39\,s. We analyse data that were reconstructed using wave-front errors obtained from the in-flight phase-diversity measurements, via an image doubler in front of the CCD camera \citep[level 3 data, see][]{hirz10, hirz11}. The SuFI data were corrected for stray light by deconvolving them with the stray light modulation transfer function (MTFs) derived from comparing the limb intensity profiles recorded for the different wavelengths, with those from the literature \citep{feller}.

\begin{turnpage}
\begin{figure*}
\centering
\includegraphics[width=\linewidth]{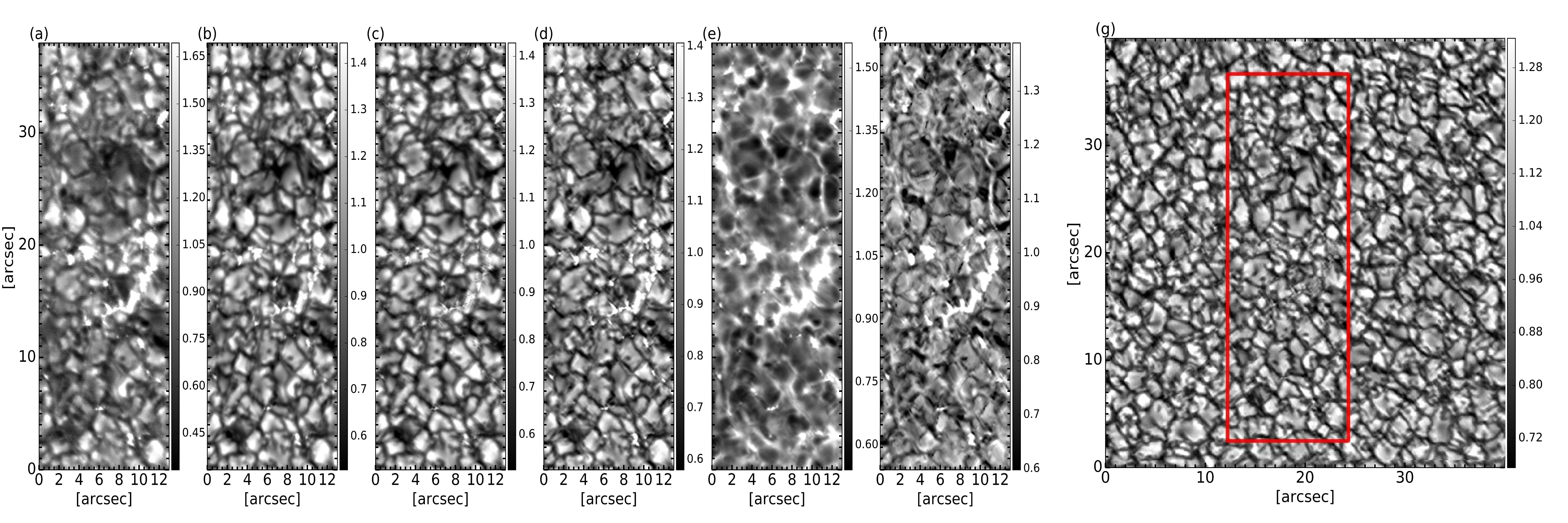}
\caption{Example of aligned SuFI and IMaX contrast images, a) SuFI at 214 nm, b) SuFI at 300 nm,  c) SuFI at 313 nm (OH-band), d) SuFI at 388 nm (CN-band), e) SuFI  Ca\,{\sc ii} H line core at 397.0 nm, f) IMaX line core, g) full FOV of IMaX Stokes~$I$ continuum at 5250.4\, \AA{}. The red box overlaid on the IMaX FOV is the common FOV ($13^{\prime\prime}\times38^{\prime\prime}$), to which the other images in this figure are trimmed. The gray scale is set to cover two times the rms range of each image.
}
\label{images}
\end{figure*}
\end{turnpage}

\begin{figure}
\centering
\includegraphics[width=\linewidth]{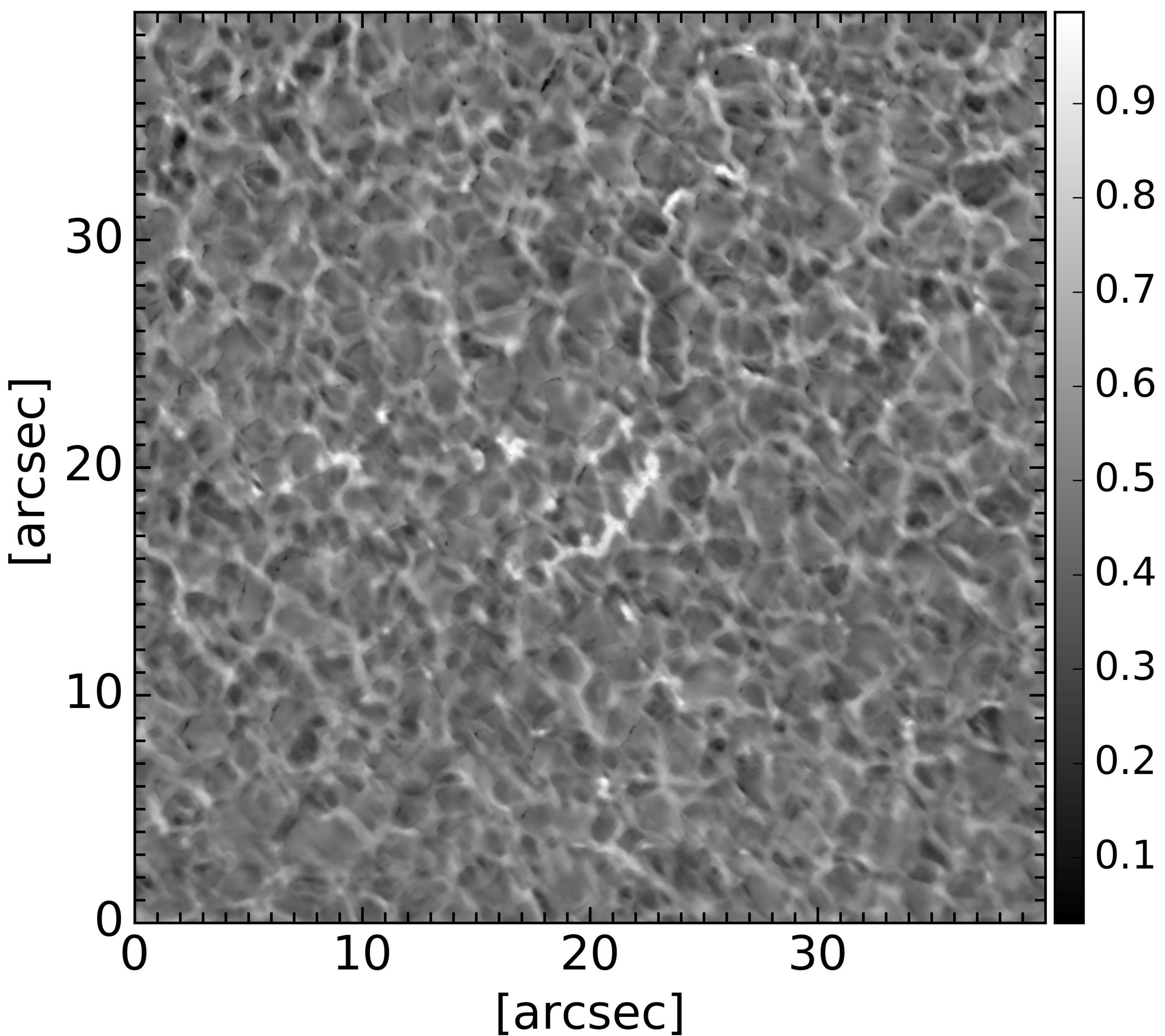}                                                                                                                                       
\caption{Same IMaX FOV shown in Fig.~\ref{images} but for Stokes~$I$ line core normalized to the local continuum.}
\label{LC_qs}
\end{figure}

\subsection{Stokes inversions}
In this study we use the line-of-sight (LOS) component of the magnetic field vector $B_{\mathrm{LOS}}$, which was retrieved from the
reconstructed Stokes images by applying the SPINOR\footnote{The {\bf S}tokes-{\bf P}rofiles-{\bf IN}version-{\bf O}-{\bf R}outines.}
inversion code \citep{Solanki1987,Frutiger2000a,Frutiger2000b}. Such an inversion code assumes that the five spectral positions within the
525.02~nm Fe\,{\sc i} line are recorded simultaneously. This is not the case for the IMaX instrument that scans through the positions sequentially
with a total acquisition time of 33\,s. To compensate for the solar evolution during this cycle time we interpolated the spectral scans with
respect to time.

We also corrected the data for stray light because we expect that the stray-light contamination has a serious effect on the inversion results,
in particular on the magnetic field results in the darker regions (intergranular lanes, micropores). Unfortunately, Feller et al. (in preparation)
could only determine the stray-light MTF for the Stokes~$I$ continuum images but not for the other spectral positions. Since a severe
wavelength dependence of the stray light cannot be ruled out, we decided to apply a simplistic global stray-light correction to the IMaX
data by subtracting 12\% (this value corresponds to the far off-limb offset determined in the continuum by Feller et al., in preparation) of the spatial mean
Stokes profile from the individual profiles.

After applying the time interpolation and the stray-light correction, the cleaned data were inverted with the traditional version of the SPINOR code.
In order to get robust results, a simple one-component atmospheric model was applied that consists of three optical depth nodes for the temperature
(at $\log\tau=-2.5, -0.9, 0$) and a height-independent magnetic field vector, line-of-sight velocity and micro-turbulence. The spectral resolution
of the instrument was considered by convolving the synthetic spectra with the spectral point-spread function of IMaX \citep[see bottom panel of
Fig.~1 in][]{Riethmueller2014}.

The SPINOR inversion code was run five times in a row with ten iterations each. The output of a run was smoothed and given as initial atmosphere
to the following run. The strength of the smoothing was gradually decreased which lowered the spatial discontinuities in the physical quantities caused
by local minima in the merit function. The final LOS velocity map was then corrected by the etalon blueshift which is an unavoidable instrumental effect
of a collimated setup \citep[see][]{mart11} and a constant velocity was removed from the map so that the spatially averaged velocity is zero.

The inversion strategy used for the 2009 IMaX data analysed in this paper is identical to the one applied to the 2013 data which is described in more detail
by \citet{Solanki2016}.

\subsection{Image Alignment}

When comparing the SuFI and IMaX data, we need to align the two data sets with each other and transform them to the same pixel scale. Because some of the SuFI wavelengths show granulation (at 300 nm, 313 nm, 388 nm, and 214 nm), these were aligned with IMaX Stokes~$I$ continuum images. The SuFI Ca\,{\sc ii} H images at 397 nm were aligned with IMaX Stokes~$I$ line-core images (see Sect.~\ref{contrast_def} for the derivation of the line-core intensity), since both wavelength bands sample higher layers in the photosphere and display reversed granulation patterns.

In a first step, the plate scale of SuFI images of roughly 0\carcsec{}02 pixel$^{-1}$ were resampled via bi-linear interpolation to the plate scale of IMaX (0\carcsec{}05 pixel$^{-1}$).

After setting all the images to the same pixel size, IMaX images with $50^{\prime\prime}\times50^{\prime\prime}$ FOV were first trimmed to exclude the edges lost by apodisation. The usable IMaX FOVs of $40^{\prime\prime}\times40^{\prime\prime}$ were then flipped upside down, and trimmed to roughly match the FOV of the corresponding SuFI images of $15^{\prime\prime}\times40^{\prime\prime}$. A cross-correlation technique was used to compute the horizontal and vertical shifts with sub-pixel accuracy. After shifting, the IMaX and SuFI FOVs  were trimmed to the common FOV (CFOV) of all data sets of $13^{\prime\prime}\times38^{\prime\prime}$. This value is smaller than the FOV of individual SuFI data sets since the images taken in the different SuFI filters are slightly shifted with respect to each other due to different widths and tilt angles of the used interference filters.
Figure~\ref{images} shows from right to left, an IMaX Stokes~$I$ continuum contrast image with an effective FOV of $40^{\prime\prime}\times40^{\prime\prime}$, with the CFOV overlaid in red, an IMaX Stokes~$I$ line core contrast image trimmed to the CFOV,  and the corresponding resampled and aligned SuFI contrast images at 397 nm, 388 nm, 313 nm, 300 nm, and 214 nm (see Sect.~\ref{contrast_def} for the definition of intensity contrast).\\
The reversed granulation pattern in the line-core image is more visible if normalized to the local continuum intensity as shown in Fig.~\ref{LC_qs}.

\subsection{Contrast}
\label{contrast_def}
The relative intensity (hereafter referred to as contrast), $C_{\rm WB}$ at each pixel for each wavelength band, $\rm WB=\{\rm CONT, \rm LC, 214, 300, 313, 388, 397 \}$ is computed as follows:

\begin{equation}
C_{\rm WB} = \frac{I_{\rm WB}}{I_{\rm WB, QS}}
\end{equation}

Where $C_{\rm CONT}$ and $C_{\rm LC}$ are the IMaX Stokes~$I$ continuum and line-core intensity contrasts, respectively. $C_{\rm 214}$, $C_{\rm 300}$, $C_{\rm 313}$, $C_{\rm 388}$ and $C_{\rm 397}$ are the SuFI intensity contrasts at 214 nm, 300 nm, 313 nm, 388 nm and 397 nm, respectively.\\

$I_{\rm WB, QS}$ is the mean quiet-Sun intensity averaged over the entire common FOV. When comparing SuFI contrast with IMaX-based magnetic field parameters this common FOV is $13^{\prime\prime}\times38^{\prime\prime}$, when comparing IMaX continuum or line-core intensity with IMaX magnetic field, the full usable IMaX FOV of $40^{\prime\prime}\times40^{\prime\prime}$ is employed.\\

The line-core intensity, $I_{\rm LC}$ at each pixel was computed from a Gaussian fit to the 4 inner wavelength points of the Stokes~$I$ profile. For comparison later in Sect.~\ref{scatter_ca}, we also compute the line core  by averaging the IMaX Stokes~$I$ intensity at $-40$ m${\angstrom}$ and $+40$ m${\angstrom}$ from the line centre:
\begin{equation}
I_{\rm LC,\pm 40} = \frac{I_{+40}+ I_{-40}}{2}
\label{LC_40}
\end{equation}

Each scatterplot in Sect.~\ref{plots} contains data from all the 40 available images in the time series.

\section{Results}
\label{plots}
\subsection{Scatterplots of IMaX continuum and line core contrasts vs. $B_{\rm LOS}$}
\label{plots_b_c}
\begin{figure}
\centering
\includegraphics[width=\linewidth]{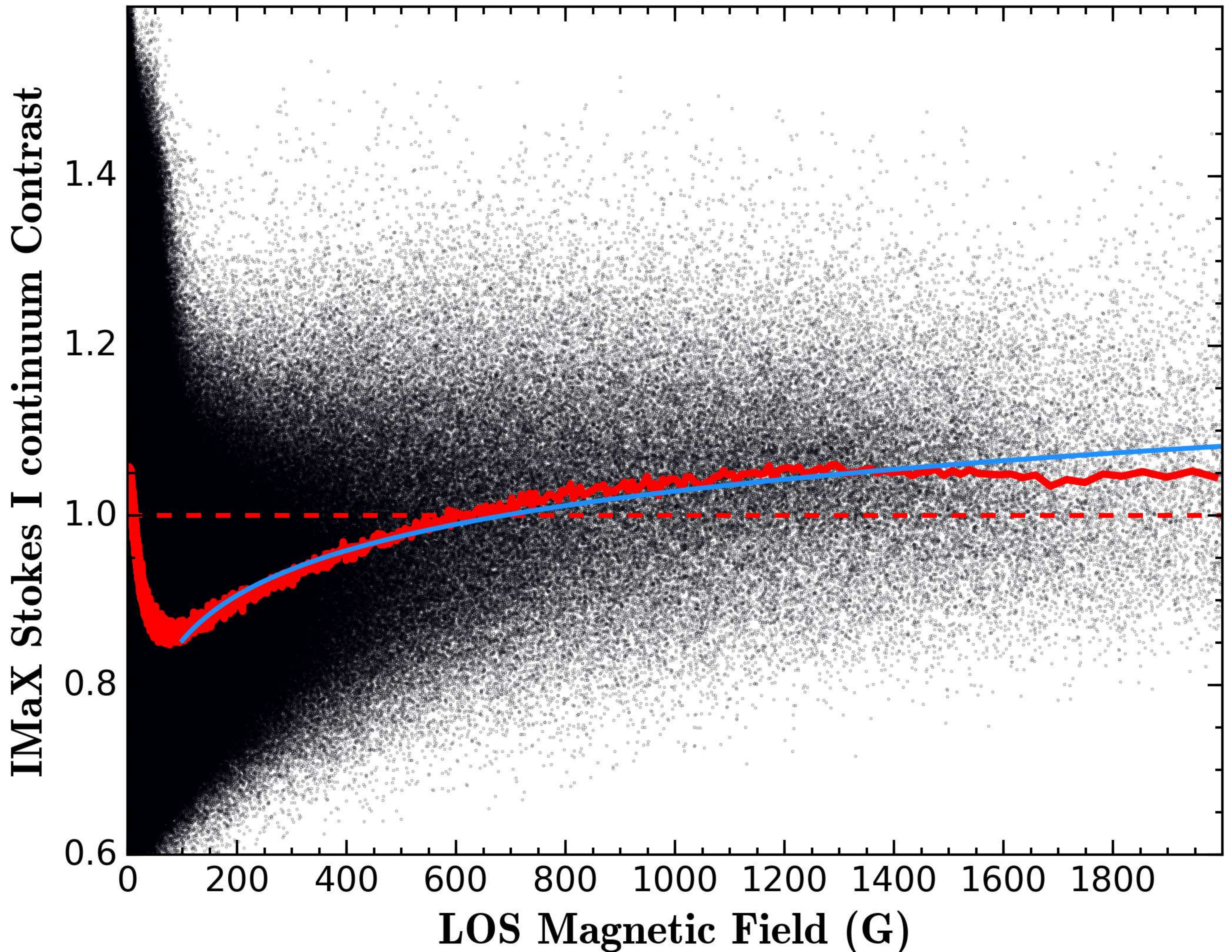}
\caption{Scatterplot of the IMaX continuum contrast at 5250.4\, \AA{} vs. the LOS component of the magnetic field in the QS at disc centre. The horizontal dashed red line indicates the mean quiet-Sun continuum intensity level, i.e. a contrast of unity. The red curve is composed of the binned values of the contrast, with each bin containing 500 data points. The blue curve is the logarithmic fit to the binned values starting at 90\,G . }
\label{cont_con}
\end{figure}

\begin{figure}
\centering
\includegraphics[width=\linewidth]{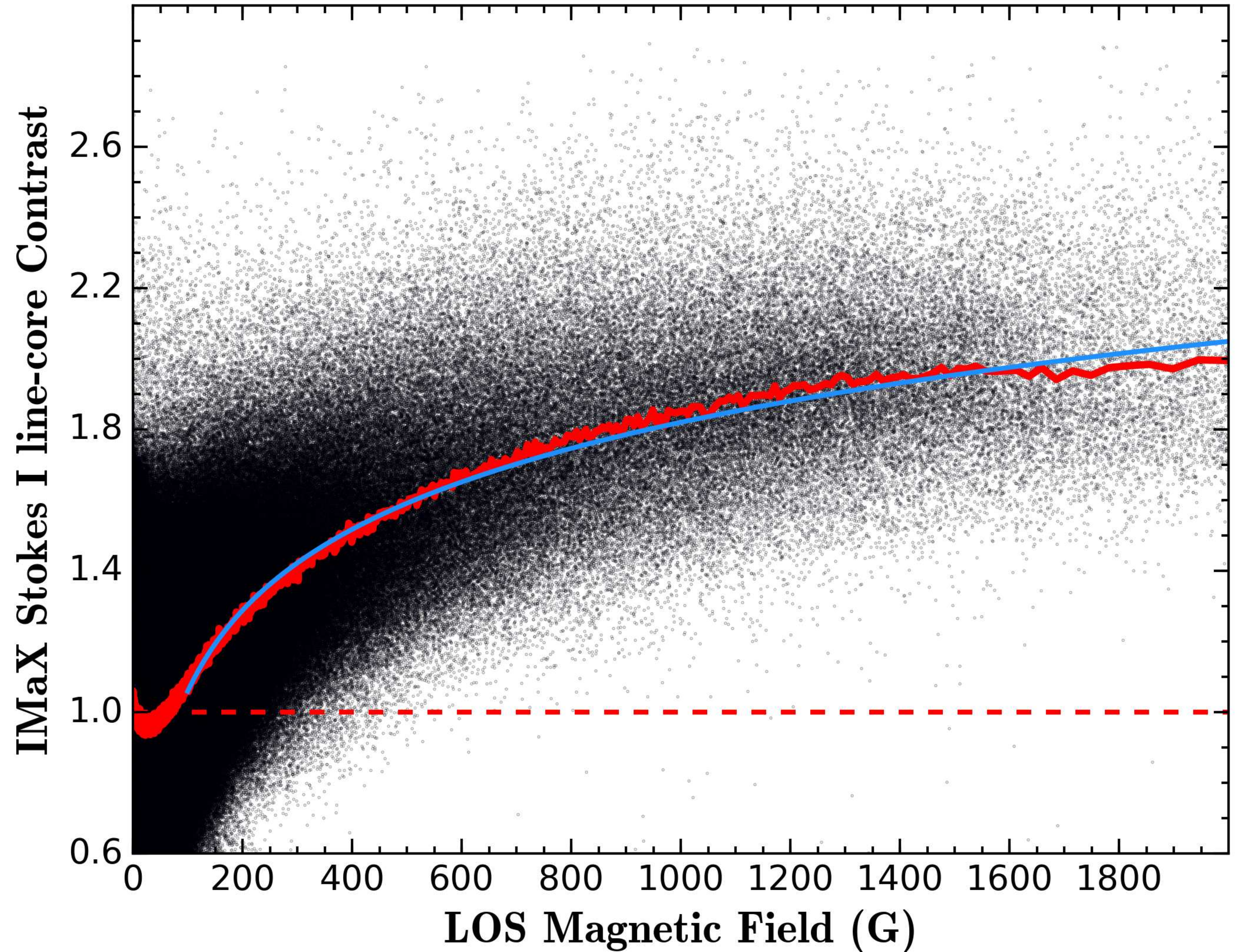}
\caption{Same as Fig.~\ref{cont_con}, but for the IMaX line-core contrast derived from Gaussian fits to the Stokes~$I$ line profiles.}
\label{lc_con}
\end{figure}

\begin{figure}
\centering
\includegraphics[width=\linewidth]{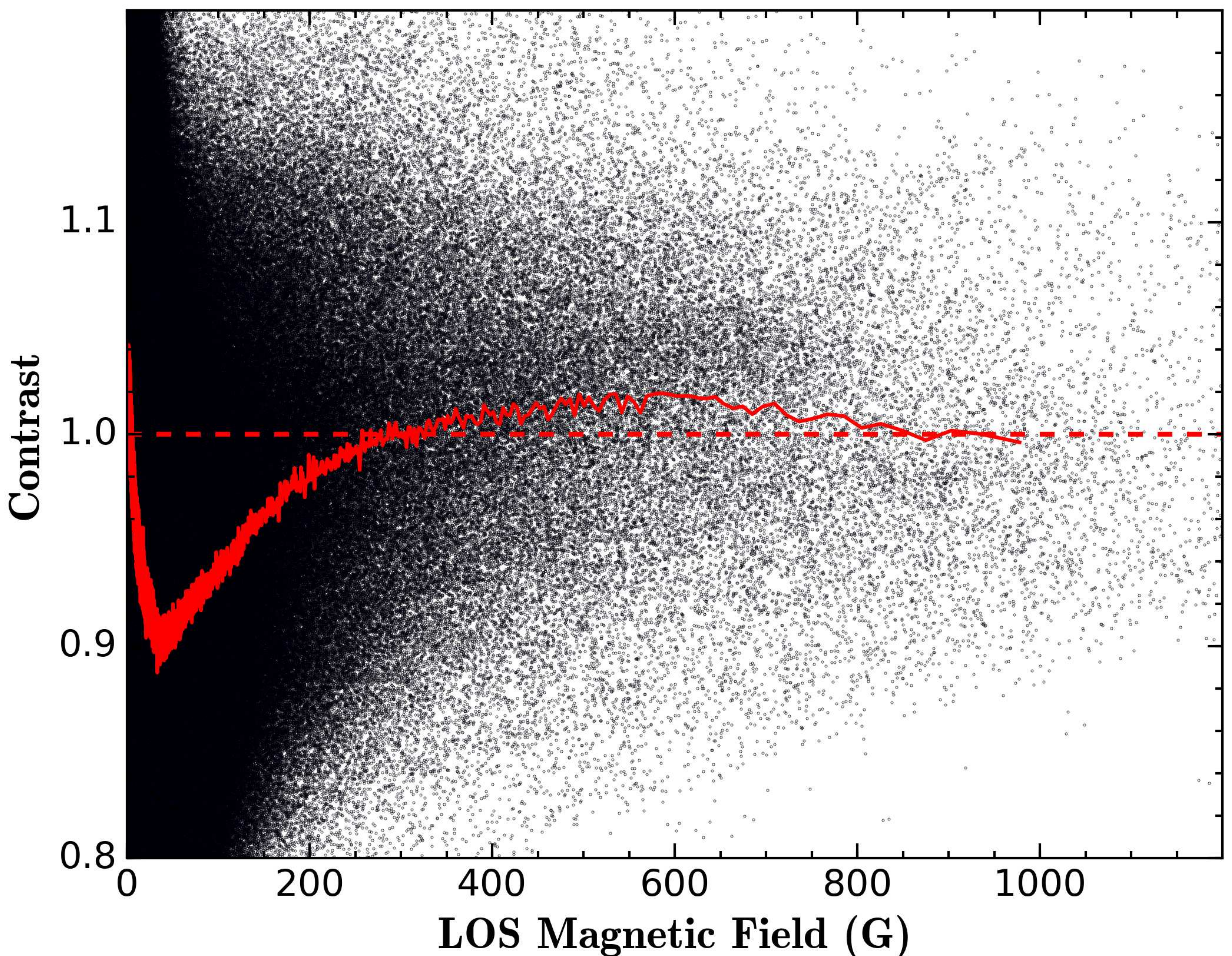}
\caption{Same as Fig.~\ref{cont_con}, but after degrading the underlying continuum Stokes images with a Gaussian of FWHM 0\carcsec{}32 to mimic the spatial resolution of the Hinode Spectropolarimter.}
\label{b_c_hinode}
\end{figure}

Pixel-by-pixel scatterplots of the IMaX continuum and line core contrasts are plotted vs. the longitudinal component of the magnetic field, $B_{\rm LOS}$ in Figs.~\ref{cont_con} and \ref{lc_con}.\\
The contrast values are averaged into bins, each containing 500 data points, which are overplotted in red on Figs.~\ref{cont_con} and \ref{lc_con}, as well as on Figs.~\ref{b_c_hinode} and \ref{sufi} that are discussed later.

For the continuum contrast (Fig.~\ref{cont_con}), the large scatter around $B_{\rm LOS} \sim 0$ is due to the granulation. At very weak fields, the average contrast decreases with increasing field strength, because only the weakest fields are present in granules, while slightly stronger fields, close to the equipartition value, are concentrated by flux expulsion in the dark intergranular lanes \citep{parker63}. These weak fields are typically at or below the equipartition field strength of around 200--400\,G at the solar surface \citep[e.g.,][]{solanki96}, which corresponds roughly to 120--240\,km about 1 scale height above the solar surface, a very rough estimate of the height at which Fe\,{\sc i} 525.02 nm senses the magnetic field. Fields of this strength have little effect on the contrast, so that these pixels are darker than the mean quiet-Sun intensity (shown as the horizontal dashed red line). The contrast reaches a minimum at approximately 80\,G, then increases with increasing field strength, as the pressure in the flux tubes decreases, and these brighten, becoming brighter than the mean QS at around 600\,G. Together, these various effects give rise to the ``fishhook" shape of the continuum contrast curve, as described by \citet{schnerr11}. The contrast then saturates at larger field strengths.

\begin{deluxetable}{cccc}[!htb]
\tabletypesize{\scriptsize}
\tablecaption{Parameters of logarithmic fits according to Eq.~\ref{log} to the continuum contrast at 5250.4\, \AA{} vs. the $B_{\rm LOS}$.}
\tablehead{
\colhead{Threshold (\,G)} & \colhead{$\alpha$}  & \colhead{$\beta$} & \colhead{$\chi^2$} \\
}
\startdata
90 &0.17 $\pm$0.001 & 0.51$\pm$0.002 & 7.62 \\
130 & 0.18 $\pm$0.001&0.47$\pm$0.002&4.04\\
170 & 0.19$\pm$0.001&0.46$\pm$0.003&3.45\\
210 & 0.19$\pm$ 0.002&0.46$\pm$0.004&3.26\\
250 & 0.18$\pm$0.002&0.47$\pm$0.006&3.05\\
\enddata
\label{cont_log}
\end{deluxetable}

\begin{deluxetable}{cccc}[!htb]
\tabletypesize{\scriptsize}
\tablecaption{Parameters of logarithmic fits according to Eq.~\ref{log} to the IMaX line core contrast vs. the $B_{\rm LOS}$.}
\tablewidth{0pt}
\tablehead{
\colhead{Threshold (\,G)} & \colhead{$\alpha$}  & \colhead{$\beta$} & \colhead{$\chi^2$} \\
}
\startdata
100 &0.76$\pm$0.002 & -0.47$\pm$0.004 & 6.3 \\
140 & 0.80 $\pm$0.002&-0.55$\pm$0.002&3.15\\
180 & 0.80$\pm$0.002&-0.58$\pm$0.007&2.55\\
220 & 0.81$\pm$ 0.003&-0.60$\pm$0.009&2.28\\
240 & 0.81$\pm$0.004&-0.60$\pm$0.01&2.22\\
\enddata
\label{lc_log}
\end{deluxetable}

The contrast values reached in the line core data (Fig.~\ref{lc_con}) are much higher than those in the continuum, in agreement with \citet{title92} and \citet{yeo13}, and are on average larger than the mean QS intensity for $B_{\rm LOS}>$ 50\,G. 

The high average contrast values (both in the continuum and line core) with respect to the mean QS for strong magnetic fields proves the enhanced brightness property of small scale magnetic elements present in our data. Moreover, the average continuum contrast values reported here, are higher than the ones measured by \citet{kobel11}, partly due to our higher spatial resolution, but partly also due to the shorter wavelength of 525 nm vs. 630 nm of the Hinode data employed by \citet{kobel11}. In addition, our plots do not show any peak in the contrast at intermediate field strengths, nor a downturn at higher values as reported by \citet{kobel11} and \citet{lawrence93}, or a monotonic decrease as obtained by \citet{topka92}.\\

To reproduce the scatterplot obtained by \citet{kobel11}, and to show the effect of spatial resolution on the shape of the $C_{\rm CONT}$ vs. $B_{\rm LOS}$ scatterplot, we degrade our data (Stokes~$I$ and $V$ images) to the spatial resolution of Hinode,  with a Gaussian of 0\carcsec{}32 \textit{FWHM}. Then, the magnetic field is computed at each pixel, using the centre of gravity (COG) technique \citep{rees79} for the degraded Stokes images.
Figure~\ref{b_c_hinode} shows the corresponding scatterplot, based on the degraded contrast and magnetic field images, with the data points binned in the same manner as for the undegraded data. One can clearly see both, a decrease in the contrast values and a leftward shift of data points towards lower magnetogram signals. Upon averaging, the binned contrast peaks at intermediate field stengths, and turns downwards at higher values.\\

To derive a quantitative relationship between the continuum contrast and the quiet-Sun magnetic field, we tried fitting the scatterplot with a power-law function of the form: 

\begin{equation}
I(B) = I_0 + a B^b
\label{pl}
\end{equation}

This function could not represent our scatterplots since the log--log plots (not shown) did not display a straight line in the range of $B_{\rm LOS}$ where we expected the fit to work, i.e. above the minimum point of the fishhook shape described earlier. Consequently, we looked for other fitting functions, and we found that the scatterplots could be succesfully fitted with a logarithmic function, which is the first time that it is used to describe such contrast curves:
\begin{equation}
I(B) = \beta + \alpha \log B 
\label{log}
\end{equation}

The fit represents the data quite well (the lin--log plots show a straight line) for data points lying above 90\,G for the continuum contrast vs $B_{\rm LOS}$, and from 100\,G for the line-core contrast vs. $B_{\rm LOS}$.
To investigate how the quality of the fit and the best-fit parameters depend on the magnetic field threshold below which all the data points are ignored, we list in Tables~\ref{cont_log} and \ref{lc_log} the best-fit parameters for the continuum and line core contrast variation with $B_{\rm LOS}$, respectively, along with the magnetic flux threshold, and the corresponding $\chi^2$ values. Fitting the original data points or the binned values returns similar results. We have plotted and tabulated the curves obtained by fitting the binned values. The $\chi^2$ values are large for smaller thresholds, and decrease with increasing threshold where less points are fitted. In contrast to this, the best-fit parameters show only a rather small variation with the threshold used, which is not the case with the power-law fit used later in Sec.~\ref{scatter_ca} when describing the relationship between the Ca\,{\sc ii} H emission and $B_{\rm LOS}$, a relationship that has traditionally been described with a power-law function.\\

In order to test the validity of the binning method used to represent the trend of the scatterplots throughout the paper, and to which the parametric models described above (logarithmic and power-law models) are fitted, we also apply non-paramteric regression (NPR) methods to the data points. These methods do not require specific assumptions about how the data should behave, and are used to find a non-linear relationship between the contrast and magnetic field by estimating locally the contrast value at each $B_{\rm LOS}$, depending on the neighbouring data points.

\begin{figure}
\centering
\includegraphics[width=\linewidth]{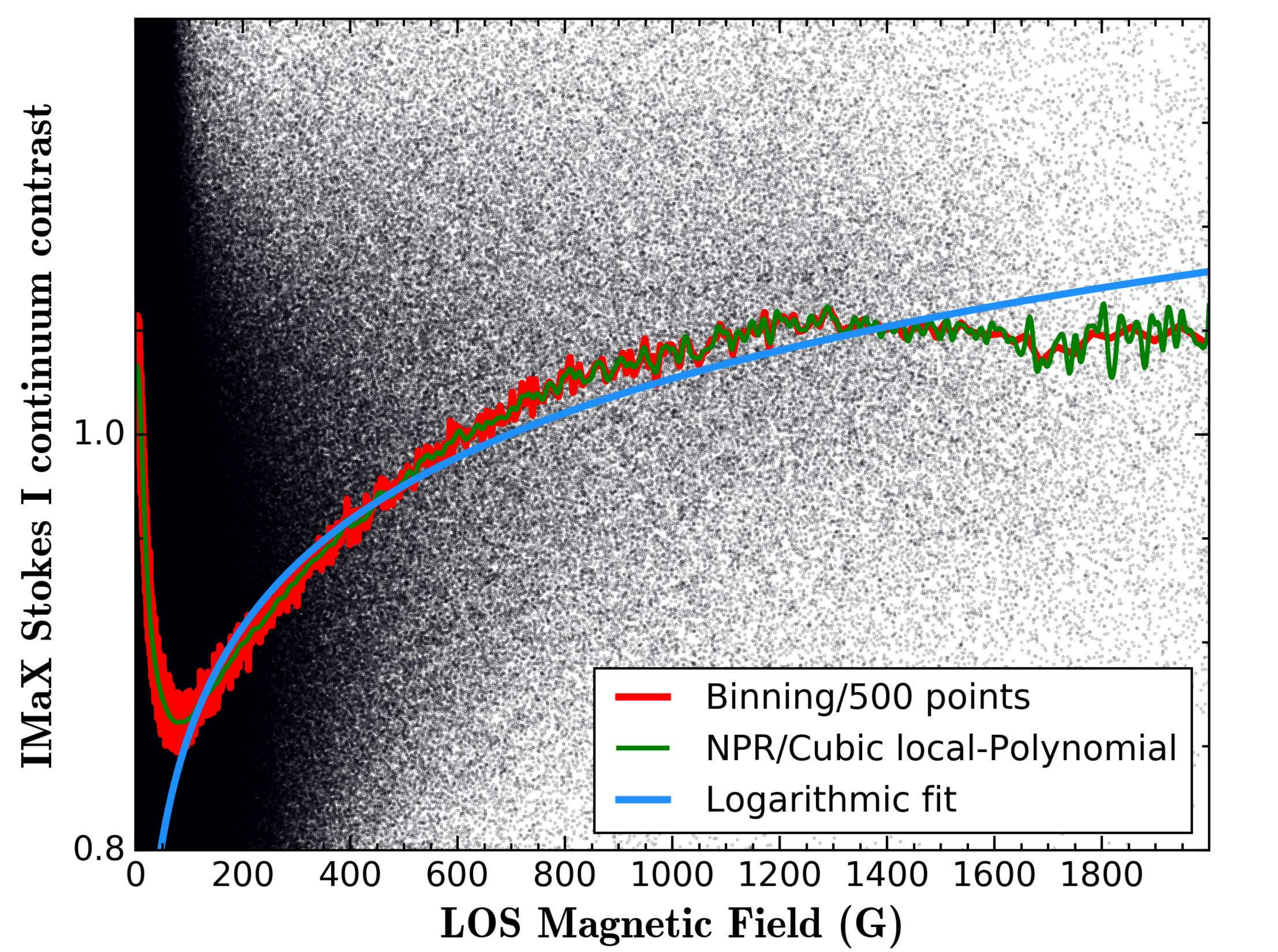}
\caption{Same scatterplot shown in Fig.~\ref{cont_con} of the IMaX continuum vs. $B_{\rm LOS}$. The red curve is the binned values of the contrast. The green curve the non-parametric regression curve obtained from applying the Kernel smoothing technique, with a local-polynomial of order $q=3$ as a regression method (check Appendix for more details). The blue curve is the logarithmic fit applied to data points starting at 90\,G and extrapolated to smaller $B_{\rm LOS}$ values.}
\label{bc_smooth}
\end{figure}

We show in Fig.~\ref{bc_smooth} the scatterplot of the IMaX continuum contrast vs. $B_{\rm LOS}$ discussed earlier in this section and depicted in Fig.~\ref{cont_con}. We plot in blue the logarithmic fit extrapolated to small $B_{\rm LOS}$ values. The red curve is the graph joining the binned contrast values and the green curve  is the regression curve obtained after applying one of the NPR techniques that are described in detail in the Appendix.

The NPR curve fits the data, including the fishhook shape at small values of $B_{\rm LOS}$, and lies extremely close to the curve produced by binning contrast values. This agreement gives us considerable confidence in the binned values we have used to compare with the simple analytical model functions.
The most relevant conclusion that can be drawn from Fig.~\ref{bc_smooth} is that our binning method is appropriate to represent the behaviour of the data points, and that it is valid to fit the logarithmic or power-law models to the binned values of the data points.

This test was repeated also for the scatterplots analysed in the next sections and the results are discussed in  Appendix.~\ref{npr_tests}.

\subsection{Scatterplots of SuFI UV brightness vs. $B_{\rm LOS}$}
\label{uv_vs_B}
\begin{figure*}
\centering
\includegraphics[width=\textwidth]{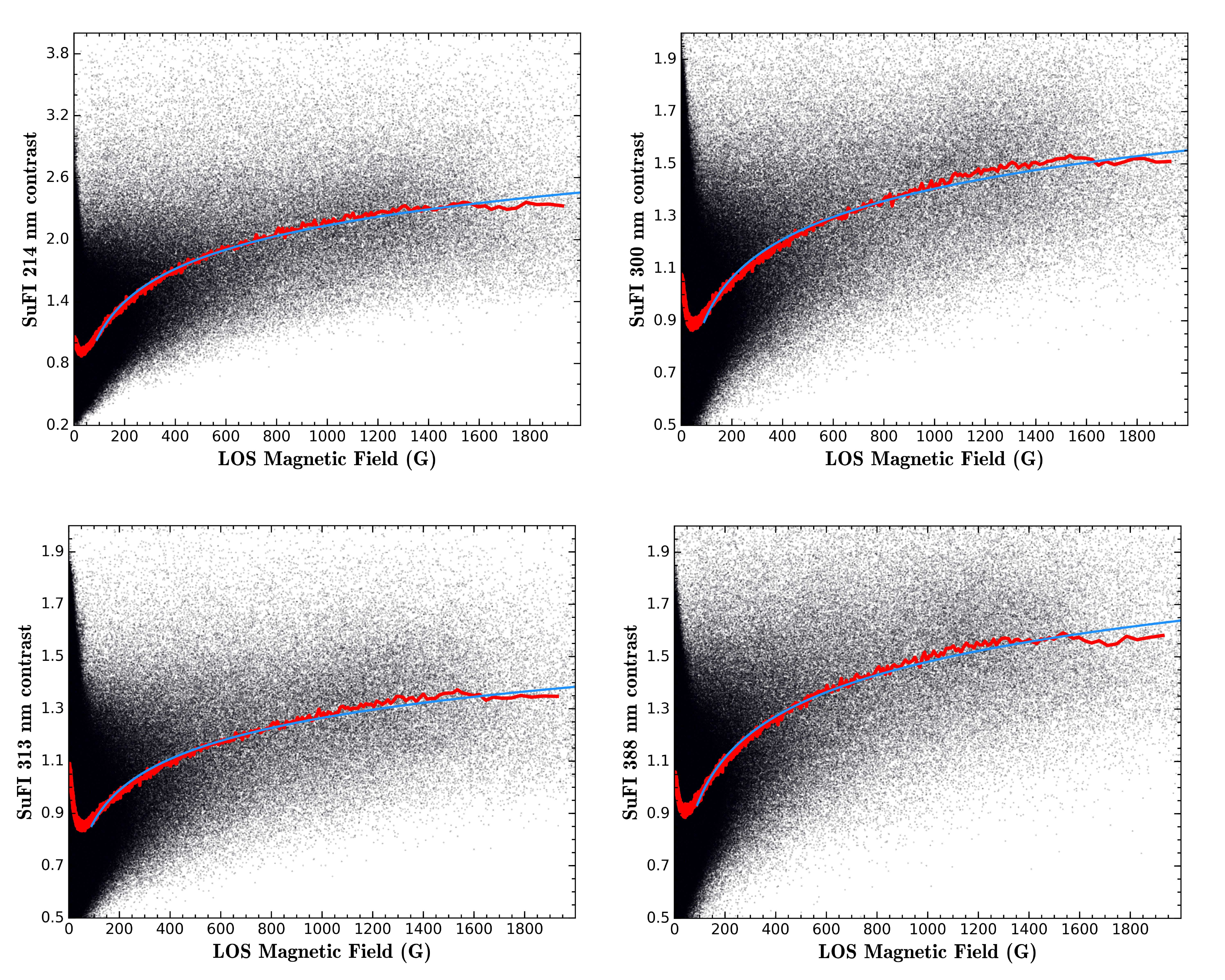}
\caption{Scatterplots of the intensity contrast relative to the average QS in four NUV wavelength bands sampled by SUFI vs. $B_{\rm LOS}$. The red curves are the values binned over 500 data points each, the blue curves are the logarithmic fits to the binned curves, starting from 90\,G,  with the fitting parameters listed in Table.~\ref{uv_log}.}
\label{sufi}
\end{figure*}

The pixel-by-pixel scatterplots of the contrast at 214 nm, 300 nm, 313 nm, and 388 nm vs. $B_{\rm LOS}$ are shown in Fig.~\ref{sufi}. The data points are binned following the procedure described in section~\ref{plots_b_c}.\\
For all wavelengths in this range, the contrasts are much larger than in the visible, especially at 214 nm where the contrast is greatly enhanced \citep[see][]{tino10}. The averaged contrast increases with increasing field strength, even for higher $B_{\rm LOS}$. Also, the fishhook shape is well visible at all UV wavelengths, and the minimum in the contrast occurs at similar $B_{\rm LOS}$ values (30\,G -- 50\,G), while contrast $>$ 1 is reached at somewhat different $B_{\rm LOS}$ values, ranging from 65\,G to 220\,G for the different UV wavelengths. 

The data points are fitted with a logarithmic function (Eq.~\ref{log}), since it describes the fitted relation better than a power-law function. Table \ref{uv_log} lists the best-fit parameters for the different spectral regions, from a threshold of 90\,G, at which the fits start to work, along with the corresponding $\chi^2$ values. 
The logarithmic function represents well the contrast vs. $B_{\rm LOS}$ relationship for all wavelengths in the NUV. A test-wise fit for different magnetic field thresholds showed that the fit results are quite insensitive to the threshold. \\

\begin{deluxetable}{cccc}[!htb]
\tabletypesize{\scriptsize}
\tablecaption{Parameters of logarithmic fits according to Eq.~\ref{log} to the observed NUV contrast vs. the $B_{\rm LOS}$. The wavelengths sampled by SuFI are shown in the first column.}
\tablehead{
\colhead{Wavelength (nm)} & \colhead{$\alpha$}  & \colhead{$\beta$} & \colhead{$\chi^2$} \\
}
\startdata

214 & 1.06$\pm$0.004 & -1.03$\pm$0.009&2.36 \\
300 & 0.48$\pm$0.002&-0.05$\pm$0.006&3.44 \\
313&0.39$\pm$0.002&0.08$\pm$0.004&2.83 \\
388&0.52$\pm$0.002&-0.08$\pm$0.005&2.54 \\
\tablecomments{The parameters correspond to logarithmic fits applied on data points fulfilling $B_{\rm LOS} >$ 90\,G}
\label{uv_log}
\end{deluxetable}

\subsection{Scatterplot of chromospheric emission vs. $B_{\rm LOS}$}
\label{scatter_ca}
\begin{figure}
\centering
\includegraphics[width=\linewidth]{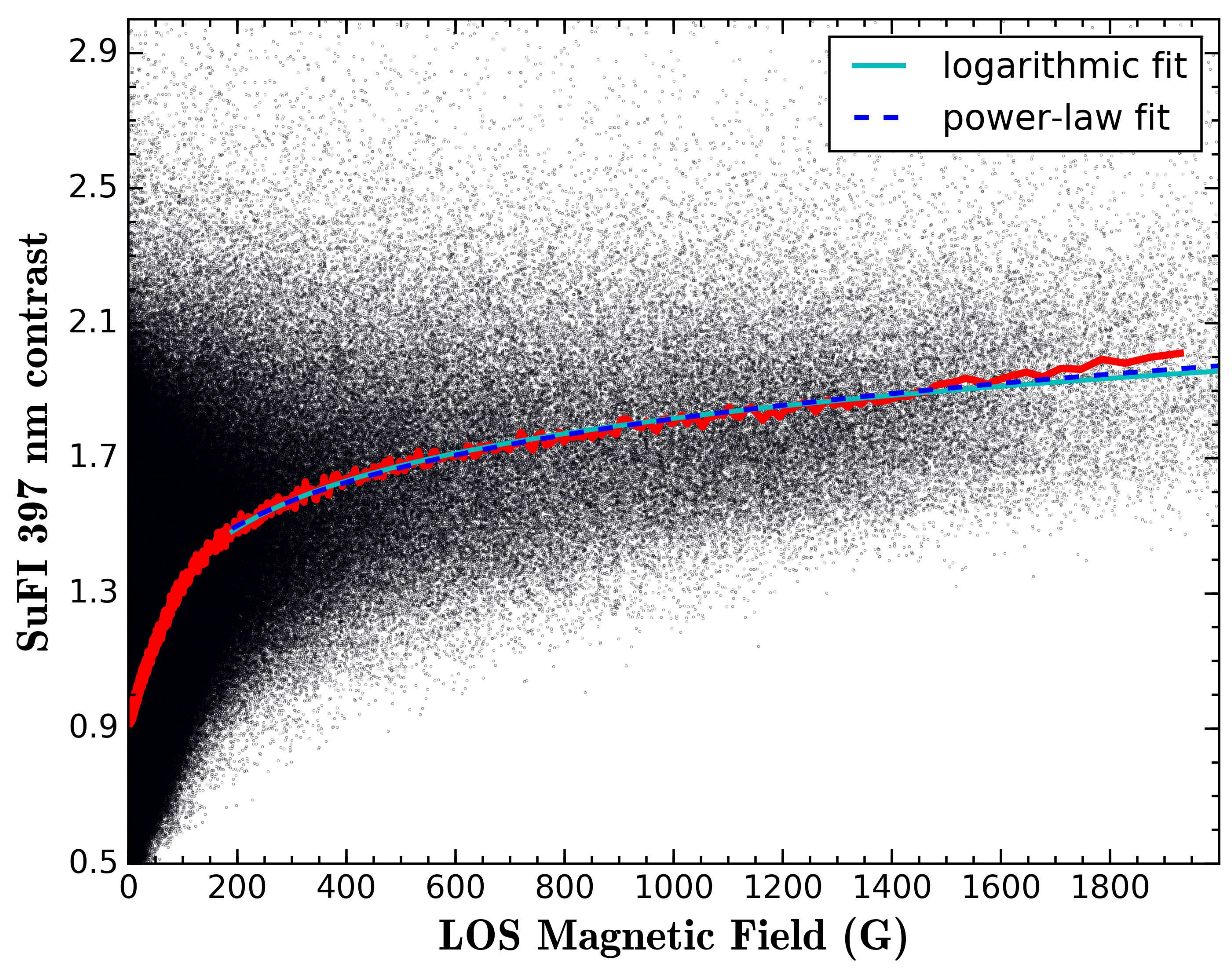}
\caption{Scatterplot of the Ca\,{\sc ii} H intensity vs. the longitudinal component of the magnetic field. The red curve represents the binned data points, the solid and dashed blue curves are the logarithmic and power-law fits to the binned data points, starting from 190\,G.}
\label{b_ca}
\end{figure}

A scatterplot of the contrast in the SuFI 397 nm Ca\,{\sc ii} H band vs. $B_{\rm LOS}$ is shown in Fig.~\ref{b_ca}.\\
The Ca\,{\sc ii} H spectral line gets considerable contribution from the lower chromosphere. This was shown in \citet{jafarzadeh13}(see their Fig.~2c), and \citet{danilovic14} (their Fig.~1) who determined the average formation heights of this line as seen through the wide and the narrow $\sunrise$/SuFI Ca\,{\sc ii} H filter, respectively, by convolving the spectra with the corresponding transmission profile and computing the contribution function for different atmospheric models. The model corresponding to an averaged quiet-Sun area returned an average formation height of 437\,km.

As pointed out in Sects.~\ref{plots_b_c} and \ref{uv_vs_B} a logarithmic function fits the contrast vs. $B_{\rm LOS}$ relationship better than a power-law function. Nonetheless, we have fitted the  Ca\,{\sc ii} H contrast vs. $B_{\rm LOS}$ relation with both, a power law (described by Eq.~\ref{pl}) and a logarithmic function (described by Eq.~\ref{log}). We first discuss the power law fits, as these have been widely used in the literature \citep[e.g.,][]{sch89,ortiz05,rezai07,louki09}. 
As pointed out by \citet{rezai07} the best-fit parameters of the power law depend significantly on the $B_{\rm LOS}$ threshold below which the fit is not applied. To investigate this dependence, we fit the data points exceeding different threshold values, which are listed in Table~\ref{ca_pl} along with the corresponding best-fit parameters and $\chi^2$ values. The data are well-represented by a power-law function, for points lying above 190\,G~\footnote{The log-log plot of the data shows a straight line starting from this value, defining where the data follows a power law function.}. However, have strongly different best-fit parameters, depending on the threshold in $B_{\rm LOS}$ applied prior to the fit.

The data were also succesfully fitted by a logarithmic function (Eq.~\ref{log}). The fit works from a lower threshold (50\,G) than the power-law fit, and the best-fit parameters vary only slightly with the threshold as can be seen in Table~\ref{ca_log}.\\
Although the log function produces a reasonable fit starting already from 50\,G, the $\chi^2$ is rather large for this threshold and drops to values close to unity only for a $B_{\rm LOS}$ threshold $>$ 190\,G, although the fit parameters remain rather stable. 
\\
\begin{deluxetable}{ccccc}[!htb]
\tabletypesize{\scriptsize}
\tablecaption{Parameters of power-law fits according to Eq.~\ref{pl} to the Ca\,{\sc ii} H emission vs. the $B_{\rm LOS}$. The first column is the threshold for the magnetic field strength, the third column is the power-law index, the fourth column is the offset, and the last one is the corresponding $\chi^2$ value.}
\tablewidth{0pt}
\tablehead{
\colhead{Threshold (\,G)} & \colhead{$a$}  & \colhead{$b$} & \colhead{$I_0$} & \colhead{$\chi^2$} \\
}
\startdata
190 & 0.61 $\pm$0.21&0.14$\pm$0.02&0.23$\pm$0.15&0.91\\
210 & 0.42$\pm$0.15&0.16$\pm$0.03&0.48$\pm$0.22&0.86\\
230 & 0.25$\pm$ 0.09&0.21$\pm$0.03&0.74$\pm$0.16&0.81\\
250 & 0.11$\pm$0.04&0.28$\pm$0.04&1.02$\pm$0.09&0.72\\
\enddata
\label{ca_pl}
\end{deluxetable}

\begin{deluxetable}{cccc}[!htb]
\tablecaption{Parameters of logarithmic fits according to Eq.~\ref{log} to the Ca\,{\sc ii} H emission vs. the $B_{\rm LOS}$}
\tablehead{
\colhead{Threshold (\,G)} & \colhead{$\alpha$}  & \colhead{$\beta$}  & \colhead{$\chi^2$} \\
}
\startdata
50 & 0.51$\pm$0.001&0.29$\pm$0.003&3.90\\
90 &0.48$\pm$0.002&0.37$\pm$0.005&1.67\\
170&0.47$\pm$0.003&0.42$\pm$0.008&1.04\\
190&0.46$\pm$0.003&0.42$\pm$0.01&1.00\\
210&0.47$\pm$0.004&0.41$\pm$0.01&0.96\\
230&0.47$\pm$0.004&0.42$\pm$0.01&0.94\\
250&0.46$\pm$0.004&0.42$\pm$0.01&0.90\\

\enddata
\tablecaption{Fitting parameters for the power law function }

\label{ca_log}
\end{deluxetable}

\begin{figure*}
\centering
\includegraphics[width=\textwidth]{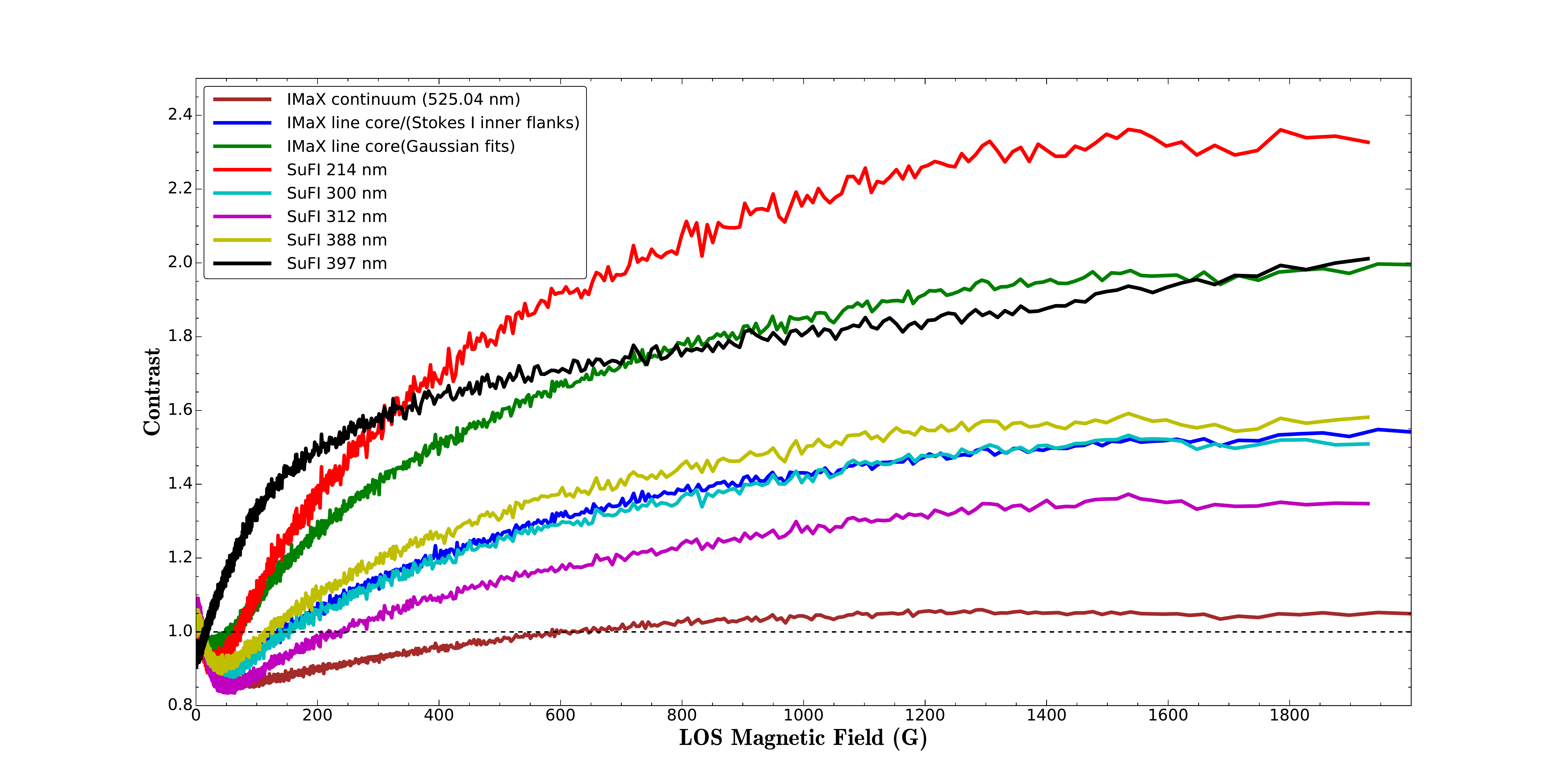}
\caption{ All binned contrast vs. $B_{\rm LOS}$ curves from Figs.~\ref{cont_con}, \ref{lc_con}, \ref{sufi} and \ref{b_ca} plotted together. Also plotted is the contrast of the 5250\, \AA{} line core obtained by averaging the intensities at the wavelength positions $+40$ and $-40$ m${\angstrom}$ apart. The curves are identified by their colour in the upper left part of the figure. The black dashed line marks the mean quiet-Sun intensity level, i.e. contrast of unity.}
\label{bins}
\end{figure*}

To better compare the magnitudes of the contrast at the different wavelengths studied here, we plot in Fig.~\ref{bins} all the scatterplots composed of the binned points obtained in the previous sections for the NUV wavelengths at 214 nm, 300 nm, 313 nm, 388 nm, and 397 nm (from Figs.~\ref{sufi} and \ref{b_ca}), and at the visible wavelengths, i.e. the continuum at 525 nm (from Fig.~\ref{cont_con}) and line core derived from the Gaussian fit to the IMaX Stokes~$I$ profiles (from Fig.~\ref{lc_con}). Also plotted is the approximate line-core value obtained from the average of the $-40$ m${\angstrom}$ and $+40$ m${\angstrom}$ line positions of the Stokes~$I$ line profile using Eq.~\ref{LC_40}.

Several qualitative observations can be made from this graph. Firstly, the contrast reached at 214 nm is higher than the Ca\,{\sc ii} H contrast. We could not find an instrumental reason for this, e.g. the calcium images were not overexposed and the response of the detector was quite linear. Therefore, we believe that the larger contrast seen in the 214 nm wavelength band is intrinsic. We expect that it is due to the very large density of lines at 214 nm and to the relatively broad Ca\,{\sc ii} H filter of 1.8 \AA{}, which has considerable contributions from the photosphere. The much higher temperature sensitivity of the Planck function at short wavelengths also plays a role.

Secondly, the large difference between the line core contrasts derived from the two different methods, with the Gaussian fits to the line profile giving almost twice the contrast (and almost reaching that of the Ca\,{\sc ii} H line core at 397 nm for higher fields) suggests that the sum of intensities at $\pm 40$ m${\angstrom}$ from the line centre is not a good approximation of the line core itself. At the other NUV wavelengths the contrast depends mainly on the number and the temperature sensitivity of the molecular lines in the passbands \citep[see][]{shussler03}. Thus the 388 nm has a larger contrast than both, 300 and 312 nm, due to the large density of CN lines in the former.

\section{Discussion and conclusions}
\subsection{Brightness in the visible vs. $B_{\rm LOS}$}

The constant continuum contrast reached in our scatterplots for field strengths  higher than 1000\,G (see Fig.~\ref{cont_con}), confirms the previous results of \citet{roh11}. They compared the relation between the bolometric intensity contrast and magnetic field strength for MHD simulations degraded to various spatial resolutions. For field strengths higher than 300--400\,G they found a monotonic increase at the original full resolution, a saturation at the spatial resolution of a 1-m telescope, while at the spatial resolution of a 50 cm telescope a turnover at around 1000\,G followed by a contrast decrease. Our results also agree with the analysis presented by \citet{kobel11}, who found that even at Hinode/SP resolution, the strong magnetic features are not resolved, leading to a turnover in their scatterplots for higher fields, similar to the behaviour obtained by \citet{lawrence93}, for a quiet-Sun region. The $\sunrise$/IMaX data display a saturation of the contrast at its maximum value in the visible continuum, in qualitative agreement with the work of \citet{roh11}, although the wavelength range of their simulated contrast is different, which indicates the need to repeat the study of \citet{roh11}, but for the actual measured wavelength bands and to compare these results quantitavely with the $\sunrise$ data. Interestingly, after degrading the $\sunrise$/IMaX data to Hinode's spatial resolution, a peak and downturn of the contrast were reproduced (see Fig.~\ref{b_c_hinode}).

The higher spatial resolution reached by IMaX allowed us to constrain the effect of spatial resolution on the relation between continuum brightness at visible wavelengths, and the LOS component of the photospheric magnetic field. At a resolution of 0.15$^{\prime\prime}$ (twice that of Hinode/SP), magnetic elements in the quiet-Sun internetwork start to be spatially resolved \citep[see][]{lagg10}, leading to a constant and high contrast becoming visible in strong magnetic features.

\subsection{Brightness in the NUV vs. $B_{\rm LOS}$ }
The relationship between the intensity and the photopsheric magnetic field for several wavelengths in the NUV provided new insights into the quantitative relation between the two parameters. The wavelength range between 200 and 400 nm is of particular importance for the variable Sun's influence on the Earth's lower atmosphere, as the radiation at these wavelengths affects the stratospheric ozone concentration \citep[e.g.,][]{gray10, ermolli13, solanki13}. Although there is some convergence towards the level of variability of the solar irradiance at these wavelengths (Yeo et al. 2015), there is a great need for independent tests of the employed modelled spectra in the UV. Such UV data are also expected to serve as sensitive tests of MHD simulations.\\

Of the UV wavelengths imaged by $\sunrise$ only the Ca\,{\sc ii} H line (discussed below) and the CN band head at 388 nm have been observed at high resolution earlier. E.g. \citet{zak05, zak07} obtain a contrast of 1.48 in bright points with the SST, which is close to the mean contrast of 1.5 we find for large field strengths.   

Most of the studies made so far on the relationship between the Ca\,{\sc ii} H emission and the photospheric magnetic field were carried out using ground-based data and different results have been obtained concerning the form of this relation. As mentioned in the introduction, several authors fitted their data with a power-law function, obtaining power-law exponents that varied from 0.2 \citep{rezai07} to 0.66 \citep{ortiz05}.
We were also able to fit our Ca\,{\sc ii} H data with a power-law function, obtaining different exponents for different thresholds of the magnetic field strength from which the fit started (see Table~\ref{ca_pl}). A nearly equally good fit was provided by a logarithmic function, starting from lower field strengths, and showing no strong variations of best-fit parameters with the threshold (see Table~\ref{ca_log}).
Other advantages of the logarithmic fit are that it has a free parameter less than the power law fit and that it also works well and equally independently of the threshold for the other observed wavelengths (see Tables~\ref{cont_log}--\ref{uv_log}), whereas the power-law fit did not lead to reasonable results.\\

As mentioned earlier in this paper, irradiance changes from below 400 nm are the main contributors to the TSI variations over the solar cycle. The magnetic flux from small-scale magnetic elements in the QS is believed to contribute considerably to not just these changes \citep{krivova03}, but likely lie at the heart of any secular trend in irradiance variations \citep[e.g.,][]{krivova07, DE16}, which are particularly uncertain, but also particularly important for the solar influence on our climate.

This contribution depends on the size and position on the solar disc of these elements and possibly on their surroundings. Here we studied the intensity contrast of a quiet-Sun region near disc centre. Next steps include repeating such a study for MHD simulations, carrying out the same study for different heliocentric angles and extending it to active region plage, so that the results can be used to test and constrain the atmosphere models used to construct spectral solar irradiance models.

\begin{acknowledgements}
The German contribution to \sunrise{} and its reflight was funded by the
Max Planck Foundation, the Strategic Innovations Fund of the President of the
Max Planck Society (MPG), DLR, and private donations by supporting members of
the Max Planck Society, which is gratefully acknowledged. The Spanish
contribution was funded by the Ministerio de Econom\'{\i}a y Competitividad under
Projects ESP2013-47349-C6 and ESP2014-56169-C6, partially using European FEDER
funds. The HAO contribution was partly funded through NASA grant number
NNX13AE95G. This work was partly supported by the BK21 plus program through
the National Research Foundation (NRF) funded by the Ministry of Education of
Korea.
\end{acknowledgements}

\clearpage

\appendix

\section{A. Non-Parametric Regression: Kernel Smoothing}
\label{appendix}

Kernel smoothing is a non-parametric regression technique where a non-linear relationship between two quantities, in our case the LOS component of the magnetic field $B_{\rm LOS}$, and the corresponding contrast value $C$ is approximated locally. In the following we write $B$ for $B_{\rm LOS}$ for simplicity. At each field value $B_0$ we want to find a real valued function $\hat{m}_h$ to compute the corresponding contrast value, $\hat{m}_h(B_0)$ or the conditional expectation of $C$ given $B_0$, i.e. $E[C|B_0]$, which is the outcome of the smoothing procedure, and is called \textit{the smooth} \citep{tukey}. The curve joining the smooth values at each $B_0$ in our data set is called the non-parametric regression or NPR curve. To do this, the function $\hat{m}_h$ is estimated at the $n$ neighboring data points,  $\{B_i,C_i\}$, falling within a bandwidth $h$ around $B_0$, with $i$ ranging from $1$ to $n$, and weighted by a kernel density function. The latter is defined for $B_0$ as:
\begin{equation}
\hat{f}(B_0) = \frac{1}{n h} \Sigma_{i=1}^{n} \mathcal{K}(\frac{B_0-B_i}{h})
\end{equation}

Here $\mathcal{K}$ is a kernel centered at $B_0$, giving the most weight to those $B_i$ nearest $B_0$ and the least weight to points that are furthest away. The shape of the kernel is determined by the type of the used kernel (often Gaussian), and by the magnitude of the bandwidth or \textit{smoothing parameter} $h$.
\\

Following a derivation that can be found in \citet{nadaraya} and \citet{watson}, the following expression for $\hat{m}_h(B_0)$ is found:
\begin{equation}
\hat{m}_h(B_0) = \frac{n^{-1}\Sigma_{i=1}^{n} \mathcal{K}_{h}(B_0-B_i)C_i}{n^{-1} \Sigma_{i=1}^{n}\mathcal{K}_{h}(B_0-B_i)},
\label{SA}
\end{equation}

which is called the \textit{Nadaraya-Watson estimator}.
\\
\subsection{A.1. Local averaging}

For local averaging, we compute at each $B_0$ a weighted average of all the $n$ data points $\{B_i,C_i\}$ that fall within the bandwidth $h$, with $i$ ranging from $1$ to $n$. Hence, the process of kernel smoothing defines a set of weights $\{W_{hi}\}_{i=1}^{n}$ for each $B_0$ and defines the function $\hat{m}_h$ as:

\begin{equation}
\hat{m}_h(B_0) = \frac{1}{n} \Sigma_{i=1}^{n} W_{hi} (B_0) C_i \\.
\label{weights}
\end{equation}

Comparing Eq.~\ref{weights} with Eq.~\ref{SA}, the weight sequence is then defined by:

\begin{equation}
W_{hi} (B_0) = \frac{\mathcal{K}(\frac{B_0-B_i}{h})}{n^{-1} \Sigma_{i=1}^{n}\mathcal{K}(\frac{B_0-B_i}{h})}
\end{equation}

\subsection{A.2. Local polynomial smoothing}
Apart from local averaging, the second kernel regression method we tested is local-polynomial smoothing. There the set of $n$ data points $\{B_i,C_i\}$ around each field strength value $B_0$ are fit with a local polynomial of degree $q$:

\begin{equation}
\hat{m}_h(B) = a_0-a_1(B-B_0)-...-a_q\frac{(B-B_0)^q}{q!}\\.
\label{taylor}
\end{equation} 
In this case, the best-fit parameters ($a_0, a_1,...,a_q$) are computed via least-squares minimization techniques, i.e the parameters that minimize the following function: 

\begin{equation}
\Sigma_{i=1}^{n} \mathcal{K}(\frac{B_0-B_i}{h})(C_i-a_0-a_1(B_0-B_i)-...-a_q\frac{(B_0-B_i)^q}{q!})^2 \\.
\label{LPq}
\end{equation}

The smooth is the value of the fit at $B_0$, i.e. $\hat{m}_h(B_0)$, which according to Eq.~\ref{taylor} is simply $a_0$. 

This procedure is applied at each $B_0$ value we have in our data, and the curve joining the smooth values $\hat{m}_h(B_0)$ is the NPR curve.

\subsection{A.3. Tests using non-parametric regression}
\label{npr_tests}
As mentioned in the main text, these techniques are applied to the various scatterplots shown in the paper to test the validity of our binning method. In Fig.~\ref{bc_smooth} we showed the scatterplot of the IMaX continuum contrast vs. $B_{\rm LOS}$ with the NPR curve plotted in green. It should be mentioned here that using a local averaging (Eq.~\ref{SA}) or a local-polynomial fit (Eq.~\ref{taylor}) returns the same result, except at the boundaries, where the curve resulting from the local-polynomial fit is smoother. The reason is the non-equal number of data points around $B$ when the latter is close to the boundary, which leads to a bias at the boundary upon locally averaging the contrast values there, meaning that the computed average will be larger than one would get if data points were symmetrically distributed around $B_0$. Using a high order polynomial reduces this bias at the boundaries, but increases the variance. 
We have tested different orders and saw that linear, quadratic and cubic polynomials give very similar results. For all the scatterplots analysed here, we only show the cubic local-polynomial ($q=3$) regression curve with a bandwidth of 6\,G\footnote{The bandwidth is taken as the square root of the covariance matrix of the Kernel used.}.
\\

\begin{figure*}
\centering
\includegraphics[width=\textwidth]{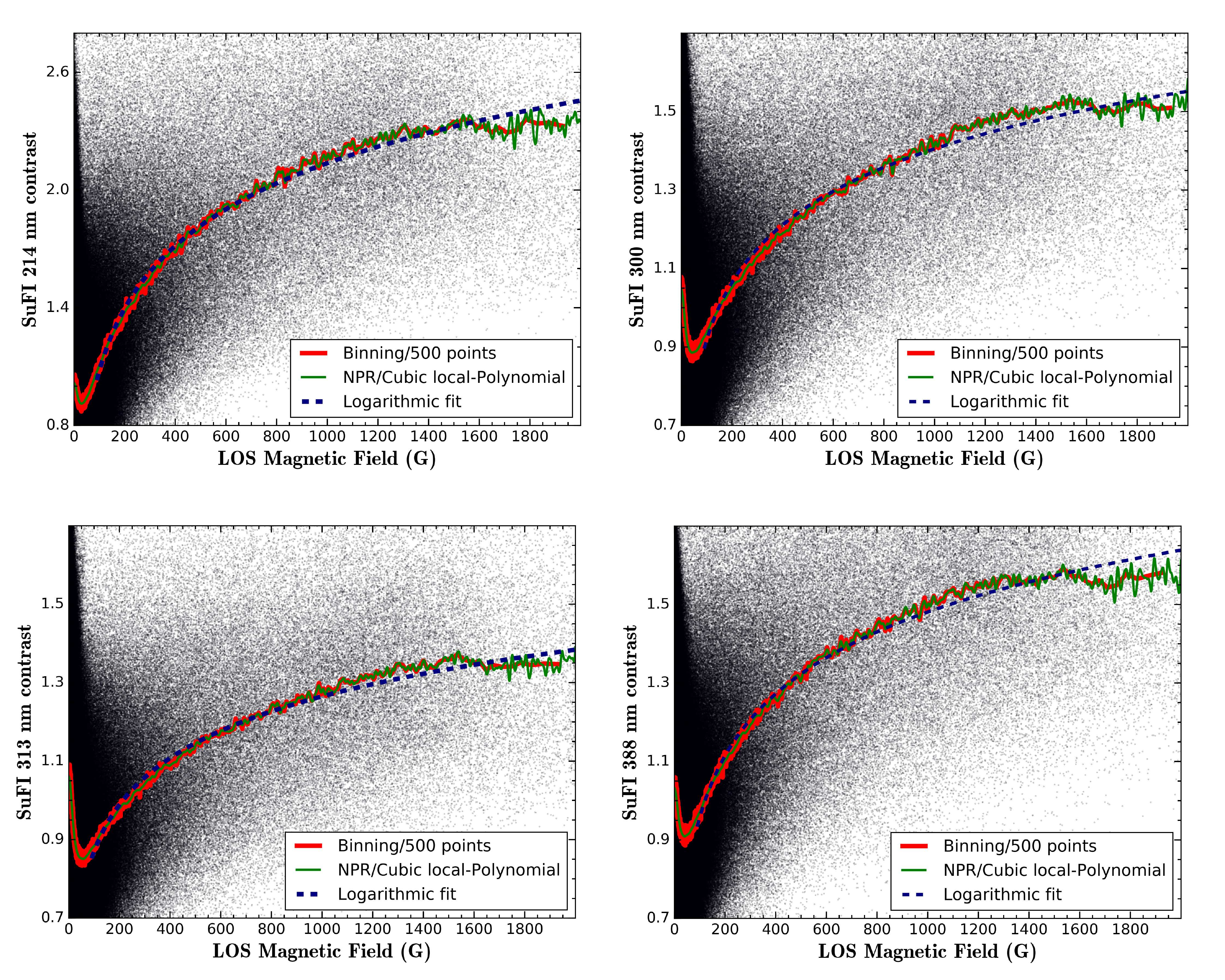}
\caption{Same scatterplots as in Fig.~\ref{sufi} of the NUV contrast vs. $B_{\rm LOS}$. The binned values are plotted in red, the NPR curves in green, and the logarithmic fits in dashed blue, starting at 90\,G.}
\label{sufi_nonparam}
\end{figure*}

Figure \ref{sufi_nonparam} shows the scatterplots analysed in Sect.~\ref{uv_vs_B}, for the NUV contrast vs. $B_{\rm LOS}$ at 214 nm, 300 nm, 313 nm, and 388 nm. The binned data points are plotted in red, the NPR curves in green, and the logarithmic fits (starting from 90\,G) in dashed blue.
The non-parametric regression curves again agree almost perfectly with the binned data (although the NPR curves are smoother, with less scatter than the binned values at small $B_{\rm LOS}$ values), and they agree very well with the logarithmic fits for almost all magnetic field values above 90\,G.\\

\begin{figure}
\centering
\includegraphics[scale=0.2]{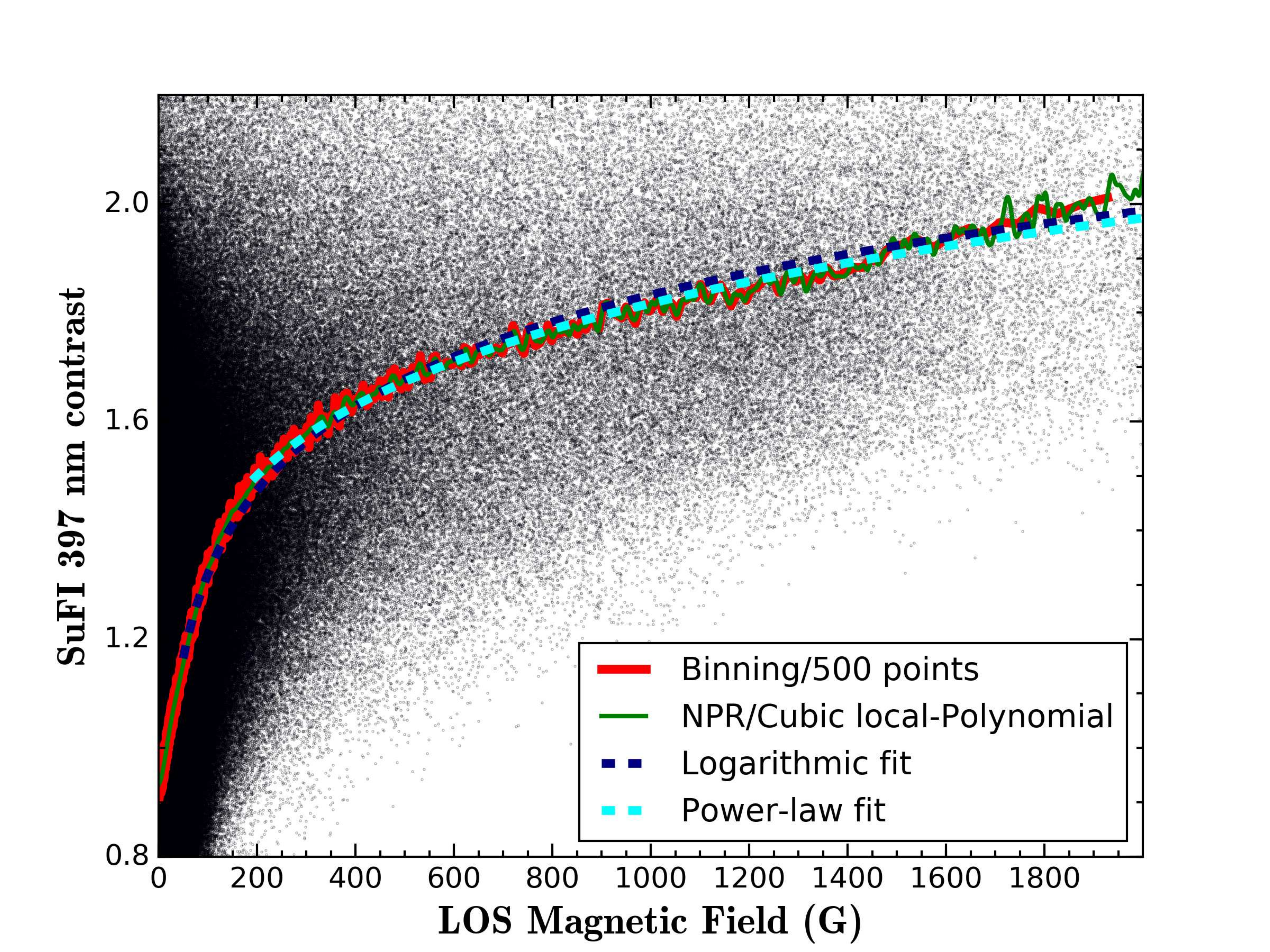}
\caption{Same scatterplot as Fig.~\ref{b_ca} of the Ca\,{\sc ii} H contrast vs. $B_{\rm LOS}$. The red curve is the binned contrast values. The NPR curve is plotted in green, with a bandwidth of $h$=6\,G, the dark blue dashed curve is the logarithmic fit to the binned curve lying above $B_{\rm LOS}$ = 50\,G, and the light blue dashed curve is the power-law fit starting at 190\,G. }
\label{b_ca_smooth}
\end{figure}

Finally, we show in Fig.~\ref{b_ca_smooth} the scatterplot of SuFI 397 nm Ca\,{\sc ii} H contrast vs. $B_{\rm LOS}$. The NPR curve is overplotted, along with both, the logarithmic fit starting at 50\,G, and the power-law fit starting at 190\,G. In addition to the similarity between the binned values plotted in red and the NPR curve (in green), it can be inferred from this plot that both parametric model fits reliably represent the variation of the Ca\,{\sc ii} H contrast vs. $B_{\rm LOS}$, since they also agree with the non-parametric regression curve for almost all values of $B_{\rm LOS}$ larger than the threshold above which the fits are applied.\\

\clearpage

\end{document}